\def\lesssim{{_ <\atop{^\sim}}}

\def\ap3m{AP$^3$M}
\def\LCDM{$\Lambda$CDM}

\def\hkpc{$h^{-1}{\ }{\rm kpc}$}
\def\hMpc{$h^{-1}{\ }{\rm Mpc}$}
\def\hMsun{$h^{-1}{\ }{\rm M_{\odot}}$}
\def\kms{${\rm{\ }km{\ }s^{-1}}$}
\def\nbody{$N$-body}
\def\c15{$c_{\rm 1/5}$}

\def\Rvir{$R_{\rm vir}$}
\def\Mvir{$M_{\rm vir}$}

\def\zform{$z_{\rm form}$}

\newcommand{\Table}[1]{Table~\ref{#1}}
\newcommand{\Sec}[1]{Section~\ref{#1}}
\newcommand{\Eq}[1]{Eq.~(\ref{#1})}
\newcommand{\Fig}[1]{Figure~\ref{#1}}
\newcommand{\mlapm}{\texttt{MLAPM}}
\newcommand{\mhf}{\texttt{MHF}}
\newcommand{\mht}{\texttt{MHT}}
\newcommand{\GKGI}{\textsf{GKG}\textrm{I}}

\def\ea{et~al.~}                            

\def\lesssim{\mathrel{\hbox{\rlap{\hbox{\lower4pt\hbox{$\sim$}}}\hbox{$<$}}}}
\def\gtrsim{\mathrel{\hbox{\rlap{\hbox{\lower4pt\hbox{$\sim$}}}\hbox{$>$}}}}

\newcommand{\ApJ}[3]    {\mbox{ApJ~\textbf{#1},~#2~(#3)}}
\newcommand{\ApJS}[3]   {\mbox{ApJ~Suppl.~\textbf{#1},~#2~(#3)}}
\newcommand{\ApJL}[3]   {\mbox{ApJ~Lett.~\textbf{#1},~#2~(#3)}}

\newcommand{\MNRAS}[3]  {\mbox{MNRAS~\textbf{#1},~#2~(#3)}}

\newcommand{\astroph}[1]{\mbox{\texttt{astro-ph/#1}}}

\documentstyle[epsfig]{mn}

\begin{document}
\title[Linking satellite dynamics to environment]
{The evolution of substructure II: linking dynamics to environment}

\author[Gill~\ea]
       {Stuart P.~D. Gill$^1$, Alexander Knebe$^1$, Brad K. Gibson$^1$,
        Michael A. Dopita$^2$\\  
        {$^1$ Centre for Astrophysics~\& Supercomputing, 
              Swinburne University, Mail \#31, P.O. Box 218, 
              Hawthorn, Victoria, 3122, Australia}\\
        {$^2$ Research School of Astronomy and Astrophysics, 
              Australian National University, Weston Creek Post Office, 
              ACT, 2611, Australia}}

\date{Received ...; accepted ...}

\maketitle

\begin{abstract}
We present results from a series of high-resolution \nbody\
simulations that focus on the formation and evolution of eight dark
matter halos, each of order a million particles within the virial
radius. We follow the time evolution of hundreds of satellite galaxies
with unprecedented time resolution, relating their physical properties
to the differing halo environmental conditions.  The self-consistent
cosmological framework in which our analysis was undertaken allows us
to explore satellite disruption within live host potentials, a natural
complement to earlier work conducted within static potentials.  Our
host halos were chosen to sample a variety of formation histories,
ages, and triaxialities; despite their obvious differences, we find
striking similarities within the associated substructure
populations. Namely, all satellite orbits follow nearly the same
eccentricity distribution with a correlation between eccentricity and
pericentre. We also find that the destruction rate of the substructure
population is nearly independent of the mass, age, and triaxiality of
the host halo.  There are, however, subtle differences in the velocity
anisotropy of the satellite distribution.  We find that the local
velocity bias at all radii is greater than unity for all halos and
this increases as we move closer to the halo centre, where it varies
from 1.1 to 1.4. For the global velocity bias we find a small but
slightly positive bias, although when we restrict the global velocity
bias calculation to satellites that have had at least one orbit, the
bias is essentially removed.
\end{abstract}


\section{Introduction}

There is mounting evidence that the Cold Dark Matter (CDM) structure
formation scenario provides the most accurate description of our
Universe. Observations point towards a ``standard'' \LCDM\ Universe
comprised of 28\% dark matter, 68\% dark energy, and luminous baryonic
matter (i.e. galaxies, stars, gas, and dust) at a mere 4\%
(cf. Spergel~\ea 2003). This so-called ``concordance model'' induces
hierarchical structure formation whereby small objects form first and
subsequently merge to form progressively larger objects (White \&
Reese 1978; Davis \ea 1985; Tormen 1997). The outcome of such mergers,
however, depends on many factors (e.g. the mass ratio of the merging
halos, their relative velocities, etc.), the result of which is a
varied mass accretion history for any given host system. While
generally successful, the
\LCDM\ model does face several problems, one such problem being the 
prediction that one-to-two orders of magnitude more satellite galaxies
should be orbiting their host halos than are observed (Klypin~\ea
1999; Moore~\ea 1999).  The lack of observational evidence for these
satellites has led to the suggestion that they are completely (or
almost completely) dark, with strongly suppressed star formation due
to the removal of gas from the small protogalaxies by the ionising
radiation from the first stars and quasars (Bullock et~al. 2000; Tully
et~al. 2001; Somerville 2002). Others suggest that perhaps low mass
satellites never formed in the predicted numbers in the first place,
indicating problems with the \LCDM\ model in general, replacing it
with Warm Dark Matter instead (Knebe~\ea 2002; Bode, Ostriker~\& Turok
2001; Colin~\ea 2000).  Recent results from (strong) lensing
statistics suggest that the predicted excess of substructure is in
fact required to reconcile some observations with theory (Dahle~\ea
2003, Dalal~\& Kochanek 2002), although this conclusion has not been
universally accepted (Sand~\ea 2003; Schechter~\& Wambsganss 2002;
Evans~\& Witt 2003). If, however, the lensing detection of halo
substructure \textit{is} correct and the overabundant satellite
population really does exist, it is imperative to understand the
orbital evolution of these objects and their deviation from the
background dark matter distribution.

The work described here focuses upon a set of numerical simulations of
structure formation within the concordance model, analysing in detail
the temporal and spatial properties of satellite galaxies residing
within host dark matter halos. To date, typical satellite properties
such as orbital parameters and mass loss under the influence of the
host halo have primarily been investigated using \textit{static}
potentials for the dark matter host halo (e.g. Johnston \ea 1996;
Hayashi
\ea 2003).  We stress that each of these studies have provided
invaluable insights into the physical processes involved in satellite
disruption; our goals was to augment these studies by relaxing the
assumption of a static host potential, in deference to the fact that
realistic dark matter halos are not necessarily axis-symmetric.  Halos
constantly grow in mass through slow accretion and violent mergers,
possessing rather triaxial shapes (Warren et~al. 1992). While a
self-consistent cosmological modeling of both hosts and satellites has
long been recognised as optimal, the required mass and force
resolution can be difficult to accommodate (hence the use of static
host potentials in most previous studies).

The first fully self-consistent simulations targeting the subject were
performed by Tormen (1997) and Tormen~\ea (1998).  Both studies were
landmark efforts, but lacked the temporal, spatial, and mass
resolution necessary to explore a wide range of environmental effects.
Unable to follow the satellite distribution within the host's virial
radius, satellites were instead tracked only up to and including the
point of ``accretion''.  This allowed an analysis of the infall
pattern, rather than the orbital evolution of the satellites.  Ghigna
\ea (1998) also investigated the dynamics of satellite galaxies in
live dark matter host halos. Although greatly increasing the mass and
spatial resolution, they still lacked the temporal resolution to
explicitly track the satellite orbits. Instead, the orbits were
approximated using a spherical static potential. More recently,
Taffoni~\ea (2003) used \nbody\ simulations coupled with
semi-analytical tools to explore the evolution of dark matter
satellites inside more massive halos. However, they focus their
efforts on the interplay between dynamical friction and tidal mass
loss in determining the final fate of the satellites.  Kravtsov~\ea
(2004) also mainly concentrate on the mass loss history of satellites
using fully self-consistent cosmological
\nbody\ simulations.

In this paper we investigate the evolution of substructure and the
orbital parameters of satellites using high spatial, mass, {\it and}
temporal resolution. As outlined in Paper~I (Gill, Knebe~\& Gibson
2004; hereafter \GKGI), our suite of simulations has the required
resolution to follow the satellites even within the very central
regions of the host potential ($\geq$5--10\% of the virial radius) and
the time resolution to resolve the satellite dynamics with excellent
accuracy ($\Delta t \approx$170~Myrs).

The outline of the paper is as follows. Section~\ref{Computation}
provides a description of the cosmological simulations employed.  The
analysis of the host halo and environment can be found in
Section~\ref{HaloAnalysis}, with the satellite orbital parameters
presented in Section~\ref{SatAnalysis}. We then investigate the
kinematic properties of the dark matter halos and satellites in
Section~\ref{velbias}. We finish with our summary and conclusions in
Section~\ref{Conclusions}.

\section{Simulation Details}\label{Computation}

Our analysis is based upon a suite of eight high-resolution \nbody\
simulations generated using the publicly available adaptive mesh
refinement code
\mlapm\ (Knebe, Green~\& Binney 2001). \mlapm\ reaches high force
resolution by refining all high-density regions with an automated
refinement algorithm.  The refinements are recursive: refined regions
can also be refined, each subsequent grid level having cells that are
half the size of the cells in the previous level.  This creates a
hierarchy of refinement meshes of different resolutions covering
regions of interest.  The refinement is done cell-by-cell (individual
cells can be refined or de-refined) and meshes are not constrained to
have a rectangular (or any other) shape. The criterion for
\mbox{(de-)refining} a cell is simply the number of particles within
that cell. A detailed study of the appropriate choice for this number
as well as more details about the particulars of the code can be found
in Knebe~\ea (2001).

The force resolution is determined by the finest refinement level
reached and corresponds to $\approx$2\hkpc\ for the simulations
presented here. The mass of an individual low-mass particle is $m_p =
1.6 \times 10^{8}$\hMsun\ and the halos are resolved with on the order
of millions of these particles.  In order to investigate the evolution
of satellite galaxies and their debris high temporal sampling of the
outputs was necessary. From $z=2.5$ to $z=0.5$ we have 17 equally
spaced outputs with $\Delta t \approx 0.35$Gyrs. From $z=0.5$ to $z=0$
we have 30 outputs spaced at $\Delta t
\approx 0.17$Gyrs.  For further details please refer to
\GKGI.

For each of our 376 outputs the satellite galaxies were initially
located using \mlapm-\texttt{H}alo-\texttt{F}inder (\mhf)
(\GKGI). This provided us with a list of all satellites and their
internal properties at each individual redshift under
consideration. However, as we are more interested in orbital
information we performed a detailed time analysis of those satellites
that were within two times the virial radius of the host halo at its
formation time (using \mlapm-Halo-Tracker: \mht). For a detailed
description of the halo finders and their effectiveness we refer the
reader to \GKGI.

\section{Host Halos and their Environment} \label{HaloAnalysis}

In \Table{HaloDetails} we summarise the relevant characteristics of
the eight host halos. The derivation of these properties is detailed
in Sections~\ref{Canon}-\ref{Supply}.

 quantities presented
discussed in this Section.

\begin{table*}
\caption{Properties of the eight host dark matter halos. Distances are measured
         in \hMpc, velocities in \kms, masses in 10$^{14}$\hMsun, and
         the ages in Gyrs.}
 \label{HaloDetails}
\begin{tabular}{cccccccccccc}\hline
halo & \Rvir & $V_{\rm circ}^{\rm max}$ & \Mvir & $T$ & $c$ &
\zform & age & $N_{\rm sat}(<\!R_{\rm vir})$ & $\sigma_{\Delta M/M}$ & formation history & richness \\ 
\hline \hline
 \# 1 &  1.34 & 1125 & 2.87 & 0.67 & 8.7 & 1.16 & 8.30 & 158 & 0.1045 & QVQQ  & 4.7\\
 \# 2 &  1.06 &  894 & 1.42 & 0.87 & 9.6 & 0.96 & 7.55 &  63 & 0.1044 & QVQQ  & 3.1\\
 \# 3 &  1.08 &  875 & 1.48 & 0.83 & 5.9 & 0.87 & 7.16 &  87 & 0.1148 & QVVVQ & 4.3\\
 \# 4 &  0.98 &  805 & 1.10 & 0.77 & 7.7 & 0.85 & 7.07 &  57 & 0.0947 & VQQ   & 2.9\\
 \# 5 &  1.35 & 1119 & 2.91 & 0.65 & 6.0 & 0.65 & 6.01 & 175 & 0.1533 & V     & 3.6\\
 \# 6 &  1.05 &  833 & 1.37 & 0.92 & 8.1 & 0.65 & 6.01 &  85 & 0.1544 & V     & 3.2\\
 \# 7 &  1.01 &  800 & 1.21 & 0.89 & 6.6 & 0.43 & 4.52 &  59 & 0.2476 & V     & 1.9\\
 \# 8 &  1.38 & 1041 & 3.08 & 0.90 & 3.7 & 0.30 & 3.42 & 251 & 0.2278 & V     & 2.8\\
\end{tabular}
\end{table*}

\subsection{Canonical Properties} \label{Canon}

A simple \mhf\ analysis of the simulation at redshift $z=0$ provides
us with the relevant information for the host halo. At $z=0$ the halo
masses range from 1--3 $ \times 10^{14}$ \hMsun\ where  mass here was
defined to be that within the virial radius \Rvir. The
virial radii in turn at which defined to be the point where the mean
averaged density of the host (measured in terms of the cosmological
background density $\rho_b$) drops below $\Delta_{\rm vir}=340$ with
\Mvir\ being the mass enclosed by that sphere. The formation
redshift \zform\ is defined here as the redshift where the halo
contains half of its present day mass (Lacey~\& Cole 1994). Applying
this criterion to our data we find that the ages of our host halos
range from 8.3 Gyrs to 3.4 Gyrs. In other words, while the masses of our
systems are comparable, they are dynamically different. In
\Table{HaloDetails}, as in all following figures, the halos are ordered
1-8 in age.

The appropriate coordinate system to investigate the orbits of the
satellite galaxies is given by the eigenvectors $\vec{E}_{\rm 1,2,3}$
(with $\vec{E}_{\rm 1}$ being the major axis) of the inertia tensor of
the respective host halo.  Moreover, the eigenvalues $a>b>c$ can be
used to describe the shape of the host and define its triaxiality
$T=(a^2-b^2)/(a^2-c^2)$ (Franx, Illingworth~\& Zeeuw 1991). The
calculation of the inertia tensor is based upon the ``core'' region of
the host as defined by the $6^{\rm th}$ refinement level in \mlapm;
i.e. the boundary of this refinement level is an isodensity
contour. The $6^{\rm th}$ refinement level surrounds material about
3000 times denser than $\rho_b$, or 9 times denser than the material
at the virial radius.  The density profiles of our host halos are well
described - at least in the range from 6\hkpc\ out to the virial
radius - by the functional form advocated by Navarro, Frenk~\& White
(1997) with concentration in the range $c\approx 4-9$. Therefore, a
density of roughly $9\times \rho(R_{\rm vir})$ corresponds to about
the half-mass radius of the host.

\subsection{Formation History} \label{FormHist}

As we are interested in investigating the influence of host halo
formation history and environment on the orbital and internal
properties of the satellites galaxies living within its virial radius,
it is important to understand how the host halos actually formed.

As indicated by Tormen (1997) the formation of dark matter halos can
be characterised by two different phases: one corresponding to a rapid
increase in halo mass, representative of a major merger. We refer to
this phase as a violent (\textbf{V}) period. The second phase is one
of relaxation in which the halo processes the merger and settles
toward virial equilibrium. A halo may continue to accrete smaller
halos during this phase. We refer to this phase as a quiet (\textbf{Q})
period. 

The history of each halo is briefly outlined below using the
simplified keys V or Q to signify violent or quiet episodes. For
example, halo~\#1 has a history ``QVQQ'', that is, a quiet period
around the time of formation $z_{\rm form}$ followed by a violent
period of merging of about the same length and completed by quiet
evolution for the remainder of the evolution, which lasts for twice as
long as either of the two earlier episodes. The coding chain splits
the evolution since formation up into more-or-less equal segments, but
has nothing to say about the absolute timescales, which differ widely
from halo to halo. A qualitative summary of the eight halos follows:

\noindent 
\textit{Halo~\#1 (QVQQ)}

\noindent
A reasonably quiet history with no major violent encounters,
save for a medium one at $z=0.7$.

\noindent
\textit{Halo~\#2 (QVQQ)}

\noindent
A generally quiet history with a medium size merger one quarter of the 
way through its evolution at $z=0.53$, quickly settling for the rest of
its formation.

\noindent 
\textit{Halo~\#3 (QVVVQ)}

\noindent
An initial short quiet period followed by a long and violent
interaction that takes essentially the rest of its formation time to
settle.

\noindent
\textit{Halo~\#4 (VQQ)}

\noindent
An initially violent merger which quickly settles for the
rest of its formation.

\noindent 
\textit{Halo~\#5 (V)}

\noindent
A very violent formation history with a strongly oscillating
potential.

\noindent
\textit{Halo~\#6 (V)}

\noindent
A steady, yet violent, formation history.

\noindent
\textit{Halo~\#7 (V)}

\noindent
A formation history quite similar to that of halo~\#6.

\noindent
\textit{Halo~\#8 (V)}

\noindent
A rapid formation history; constantly interacting
with two other large halos. In this sense, a unique
system.

While useful, such a qualitative description needs to be augmented
with a quantitative one. In order to do so, we use the dispersion of
the rate of relative mass change of the host halos:

\begin{equation}
 \sigma_{\Delta M/M}^2 = \frac{1}{N_{\rm out}} \sum_{i=2}^{N_{\rm out}}
                        \left(\frac{\Delta M_i}{\Delta t_i M_i} - 
                         \langle \frac{\Delta M}{\Delta t M}\rangle_i \right)^2 \ ,
\end{equation}

\noindent
where $N_{\rm out}$ is the number of available outputs from formation
\zform\ to redshift $z=0$, $\Delta M_i = M(z_i)-M(z_{i-1})$ the 
change in the mass of the host halo, and $\Delta t_i$ the respective
change in time. The mean growth rate at time $i$

\begin{equation}
 \langle \frac{\Delta M}{\Delta t M}\rangle_i = 
 \frac{1}{N_i} \sum_{j=1}^{N_i} \left( \frac{\Delta M_i}{\Delta t_i M_i} \right)_j
\end{equation}

\noindent
is calculated for each individual output as the average over all halos
$N_i$ available at that time step.

A large dispersion $\sigma_{\Delta M/M}$ now indicates a violent
formation history whereas low values correspond to quiescent formation
histories. As we can see in
\Table{HaloDetails}, our qualitative classification scheme is confirmed by the
$\sigma_{\Delta M/M}$ values.

\subsection{Satellite Mass History}

\begin{figure}
   \centerline{\psfig{file=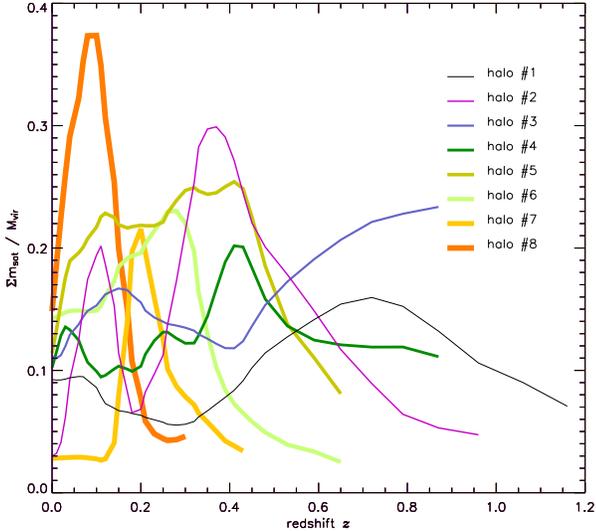,width=\hsize}}
   \caption{Mass locked-up in satellites within the virial radius
   divided by the virial mass of the host halo as a function of
   redshift.}  \label{MsatFrac}
\end{figure}

To gain further insight into the hierarchical build-up of the host and the
evolution of its satellite population we plot the fraction of its mass
locked up in the satellite distribution in \Fig{MsatFrac}.  We show the
total mass of all the satellites (identified by the \mht\ method outlined
in \GKGI) living within the virial radius divided by the mass of the host
halo as a function of redshift. 

\begin{figure}
   \centerline{\psfig{file=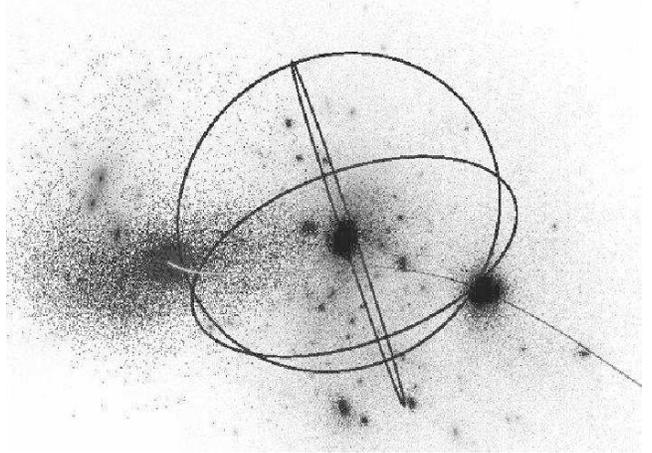,width=\hsize}}
   \caption{This figure shows a satellite of mass 10\% the host halo 
            that is responsible for the jump in the host's mass accretion history.
            The host halo (halo \# 7) is plotted as a line of sight density 
            projection using every 3$^{\rm rd}$ particle, whereas the satellite is
            represented by all its particles without a density map.  
            The satellite enters the halo at $z=0.25$ (lower right) and 
            leaves again around $z=0$ with the thin line indicating its orbit. 
            The sphere shows the virial radius of the halo at $z=0$ and
            the relative velocities of the two objects is 750\kms.}  
   \label{transat}
\end{figure}

At $z=0$ the average mass of all substructure is approximately 8--9\%
of the host halo's virial mass with the scatter allowing for as much
as 15\% and as little as 3\%. We do not observe any pronounced trend
for this fraction to depend on the age of the dark matter halo, which
is consistent with the hierarchical model of satellite accretion: both
small and large objects continuously fall in. The history is clearly
reflected in \Fig{MsatFrac}, where the infall of large satellites
gives rise to the spiky nature of the curves. These large variations
are, however, a combination of massive halos merging via dynamical
friction and ``transitory structures''.  Transitory structures are
small subsets of satellites that interact with a halo, but are not
bound to it. An example is given in
\Fig{transat} where we show one of these transitory events for halo~\#7.
The peak near redshift $z=0.2$ for this halo in
\Fig{MsatFrac} is caused by an object of roughly 10\% of the mass of
the host orbiting in the outskirts (but still within \Rvir) of the
host halo at a relative speed of approximately 750\kms.  \Fig{transat}
captures this event showing the host and its virial radius at redshift
$z=0$ with the path of the satellite indicated by a line. The
perturber itself is represented by its particles. Its tidal disruption
while passing near the host can also be appreciated in \Fig{transat}.
We particularly highlight this galactic encounter not only to explain
the rise in \Fig{MsatFrac} but also to raise the readers attention to
the potential of already ''harassed'' galaxies falling into the host
halo. The large peak for halo~\#8 in
\Fig{MsatFrac}  is due to an interaction with one of the other two
large objects in the system. The violent history of halos~\#5 and \#6
can also be seen in \Fig{MsatFrac}. The more quiescent halos also
stand out in this figure, with less variation in the substructure
evolution except for halo~\#2 with a 15\% mass merger at $z=0.53$.

In summary, we have selected a sample of halos displaying widely
different formation histories, which should aid in gaining insight
into the environmental effects of halo formation.

\subsection{The Supply of Satellites} \label{Supply}

We now investigate the temporal evolution of satellite accretion and
tidal disruption as a function of host halo environment and richness.

\begin{figure}
   \centerline{\psfig{file=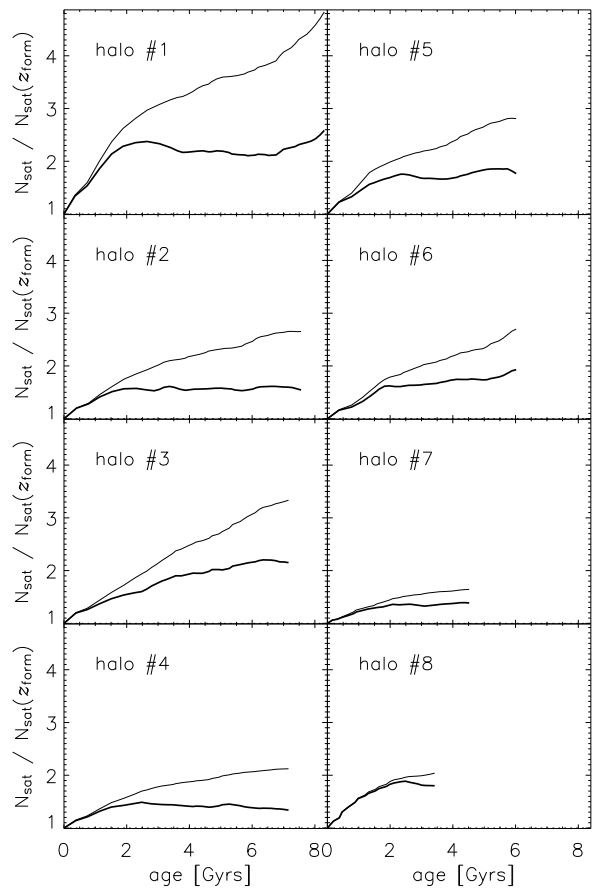,width=\hsize}}
   \caption{Number of satellites within the virial radius of their respective host 
            as a function of redshift. The thin line is the total
            number of satellites that have fallen into the host
            whereas the thick line 
            shows the number of surviving satellites. Both 
	    lines are normalised to the total number of satellites at \zform.}
   \label{NsatHist}
\end{figure}

In \Fig{NsatHist} we display the normalised number of satellites
within the respective host halo as a function of time after the
initial formation epoch. As we intend to measure ``supply rates''
rather than absolute numbers of infalling satellites we normalise the
curves by the number of satellites present at the formation time of
the host halo. The thin line represents the total normalised number of
satellites which have been accreted, while the thick line refers to
the number of \emph{surviving} satellites.  The criterion used to
define tidal disruption is the reduction in the number of particles
within a given satellite's tidal radius to fewer than 15.  This
definition is somewhat arbitrary, although ultimately based upon the
numerical resolution. For a more detailed discussion please refer to
\GKGI.

The increase in the total number of satellites (thin curve) reflects
the ``richness'' of the environment around the halo: halos with a
steep slope benefit from a constant supply of satellite galaxies
wheres hosts that only show a mild increase draw upon a pool of fewer
satellites in their immediate vicinity. This is illustrated by the
case of halo~\#1 which lies in a particularly rich environment in
which several filaments intersect (cf. \Fig{spiderpanel} below). As a
consequence, it accretes a total of nearly five times the initial
number of satellites while simultaneously showing a high satellite
disruption rate. The case of halo~\#3 is similar, but less extreme.
Halos~\#7 and \#8, by contrast, experience only moderate infall, and
in these halos nearly all the satellites survive. The situation is
illustrated for halos~\#1 and \#8 in \Fig{spiderpanel}, which shows
the orbital paths followed by all the satellites from the formation
epoch up to the present day.  In the upper panel we clearly see the
filament arms that feed halo~\#1 and how the satellites spiral into
the dark matter halo. The filaments are helical because they consist
of smaller satellites orbiting a larger host that is falling into the
massive host halo. The small but rapid rise in the satellite infall
for halo~\#1 (\Fig{NsatHist}) is caused by a group of satellites
falling into the halo for the first time. In the bottom panel of
\Fig{spiderpanel} we feature halo~\#8;  in  contrast to
halo~\#1, halo~\#8 was formed in a relatively isolated region which
saw a rapid collapse.  We can, however, confirm that even though the
satellite accretion rate in halo~\#8 is far less the mass of the
infalling objects is much higher. This is derived from the fact that
halo~\#8 acquires half its mass by digesting those few satellites in a
time span of approximately 3~Gyrs (cf. \Fig{MsatFrac}). We note though
that there also exists a significant age difference between halo~\#1
and \#8; this explains why halo~\#8 satellites are traced for a shorter time
leading to the ``shorter'' lines in
\Fig{spiderpanel}.  However, there  still exist noticeable differences in the
satellite accretion curves for halos~\#1 and \#8 (cf. \Fig{NsatHist})
when restricting halo~\#1 to the first 3.5 Gyrs of its existence.

We define the substructure ``richness'' as the ratio of the final to
initial number of satellites, and list its value for each of the eight
halos in \Table{HaloDetails}.

The number of surviving satellites is not directly correlated with the
richness, but rather to the orbital characteristics of the accreted
satellites, as we will demonstrate in detail below. In general, the
accreted satellites are not immediately disrupted, but are
progressively destroyed over time.

\begin{figure}
   \centerline{\psfig{file=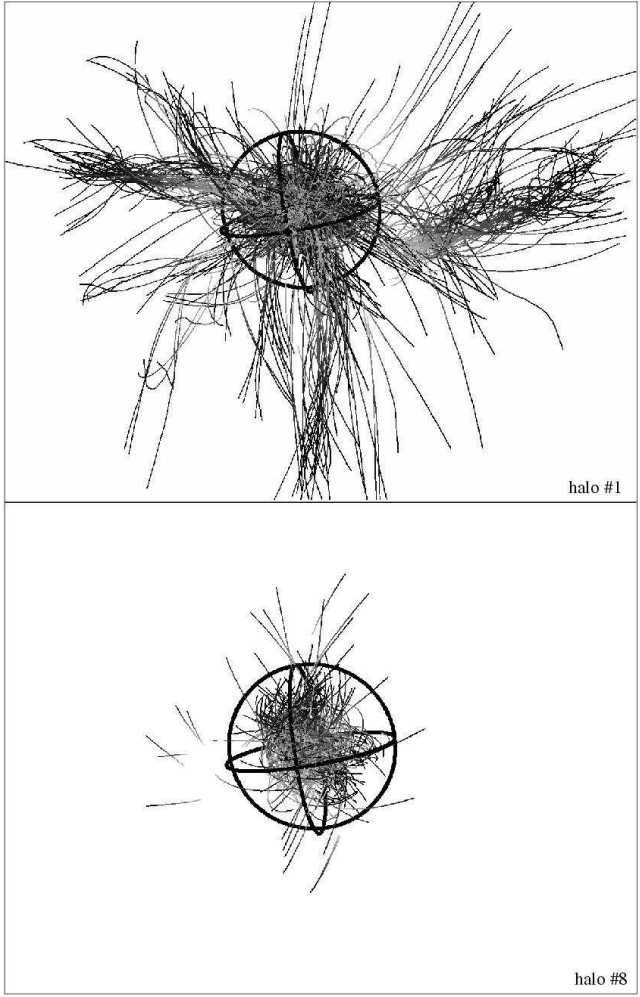,width=\hsize}}
   \caption{The orbits of all objects in the vicinity of the host halo
            are shown as lines that graduate from dark at \zform\  
            to light $z=0$. The black spheres have a 2\hMpc\ radius. The
	top panel shows the ``rich environment'' of halo \#1, the bottom
	panel the more isolated region halo \#8. While satellites 
	continually infall into halo \#1, halo \#8 saw an early, 
	rapid infall of essentially all its associated satellite substructure.}
   \label{spiderpanel}
\end{figure}

\begin{figure}
   \centerline{\psfig{file=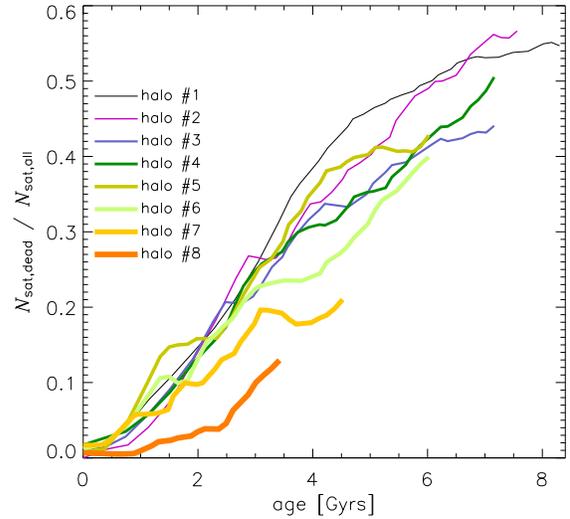,width=\hsize}}
   \caption{The ratio of  ``disrupted'' satellites to the total
            number of  satellites for
	satellites within the virial radius of their respective host, as a function
	of time, where zero age is the formation time of the halo.}
   \label{NsatHistAGE}
\end{figure}

To further investigate the link between satellite disruption and
satellite infall we calculate the ratio of disrupted (or ``dead'')
satellites to the total number of satellites that fall into the host
halo. The result is presented in \Fig{NsatHistAGE}. It is now possible
to interpret the slope of this figure as the ``rate of disruption'' of
satellite galaxies. For all halos this disruption rate (i.e. slope) is
very similar.  There seems to be no strongly pronounced correlation
with either mass, environment or age. In this respect, the destruction
rate of satellite galaxies appears to be ``common'' in CDM
halos. However, there also appears to be a (marginal) trend that host
mass is related to the ability to disrupt satellites (within the
limited mass range presented here). This can be seen explicitly for
halos~\#1 and \#3: \#1 destroys its satellites more efficiently than \#3
and is also the more massive of the two. However, these halos
also have a large difference in triaxiality (refer to
\Table{HaloDetails}). Conversely, as halo~\#3 is comparable 
in mass to halo~\#4 with a strikingly similar satellite destruction
rate. But as halo~\#3 and \#4 are close in triaxiality, we rather
suspect the disruption rate to correlate with mass than with
triaxiality. However, this correlation - if in fact correct - must
also have significant scatter, as the most massive halo (\#8) shows
the lowest destruction rate. This might be linked to the environment
and supply of new satellites, respectively.

\subsection{Summary of Host Halos and Satellite Supply}

Table~\ref{HaloDetails} summarises the basic properties of our halos
and their environments, illustrating the variety of richness and
accretion histories sampled by our simulations. The mass spectra of the
satellite galaxies (although not presented) are consistent with other
studies in the literature, being described by a declining power-law
$dn/dM\propto M^{-\alpha}$ with $\alpha \approx 1.7-1.9$ (cf.
Ghigna~\ea 2000) in the range from $2\times10^{10}$\hMsun\ (applied
mass-cut corresponding to 100 particles, which explains the rather
'low' number for $N_{\rm sat}(<\!R_{\rm vir})$ in \Table{HaloDetails})
up to $\sim10^{13}$\hMsun.  However, the number of satellites changes,
because a) new ones are constantly being accreted (cf. thin lines in
\Fig{NsatHist}) and b) they are being concurrently disrupted 
(cf. \Fig{NsatHistAGE}).  We also observe that the destruction rate is
nearly identical for all our host halos, making it (nearly)
independent of mass, age, and triaxiality.  Further, as seen in
\Fig{spiderpanel} satellites do not fall in isotropically (Knebe~\ea
2004). Instead they are accreted via large-scale filamentary
structure, with some carried in by more massive satellites with their
own substructure field, confirming the results from earlier lower
resolution simulations (Hatton~\& Ninin 2001; Colberg~\ea 1999; Tormen
1997)

\section{Orbital Parameter Statistics}\label{SatAnalysis}

Having described the differences in the host halos, we now turn to the
detailed analysis of the evolution of satellite galaxies. In the
following subsections we investigate their lifetimes, orbital
parameters, and derive relevant correlations and relations.

\begin{figure}
\centerline{\psfig{file=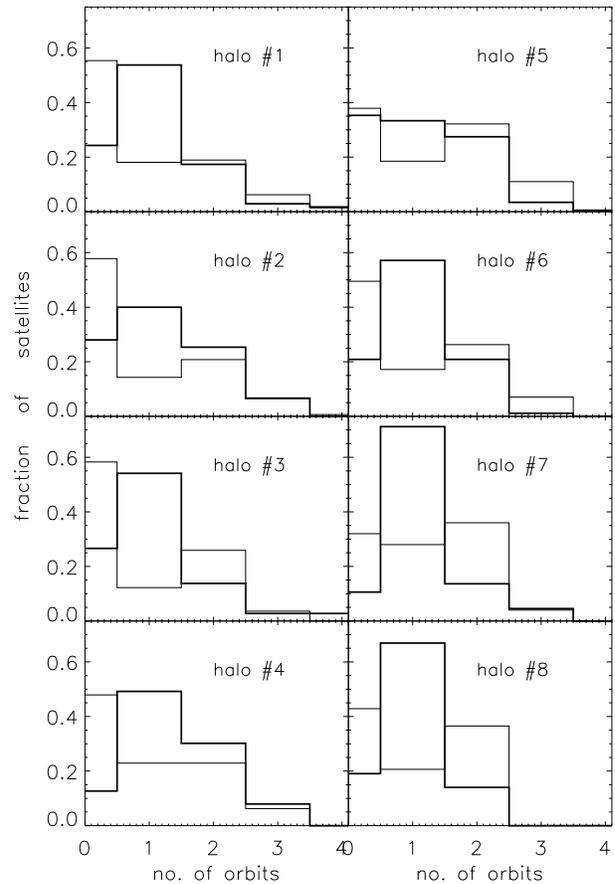,width=\hsize}}
\caption{The thick histograms show the distribution of the number of 
         orbits for all satellites that survive until redshift $z=0$. 
         The thin histograms show the distribution for disrupted satellites.}  
\label{Norbit}
\end{figure}

\subsection{Number of Satellite Orbits} \label{satOrbs}

In \Fig{Norbit} we show the distribution of the number of orbits for
both surviving satellites (thick lines) and disrupted satellites (thin
lines) at redshift $z=0$. For the determination and definition of the
number of orbits refer to \GKGI.

For the satellites that survived until $z=0$ the distribution peaks at
or near one orbit for each of the eight host halos.  More than 70\% of
the satellites have had at least one full orbit with some having as
many as four, therefore making a study of the satellite dynamics
valid. The length of the tail to the right of the peak at one orbit is
somewhat correlated with the age of the dark matter halo. Further, the
peak is more pronounced in the younger halos.  However, halos~\#1 and
\#3 both  show distinctive peaks at one orbit which  relates to
the richness of environment given in \Table{HaloDetails}, rather than
the age, as satellites are continuously falling in.

The interpretation of the distribution for the disrupted satellites is
more interesting. We note that these orbits are determined by
mass-less tracer particles placed at the last credible centre of the
satellite before disruption. For our definition of ``disruption''
please refer to \GKGI. The distributions are generally ``flatter''
with the most prominent peak near zero orbits. One explanation for
such a distribution is that the infalling satellites are being
disrupted before completing one orbit. This suggests that satellites
contributing to this bin in the distribution are very massive and
rapidly decay via dynamical friction. Having said that, we emphasize
that this interpretation may be somewhat simplistic. All satellites
were identified at
\zform\ and we do not have an indication as to how long the satellites had
already existed within the host's progenitor. Therefore, (massive)
satellites that had already been orbiting within the progenitor would
be the first to be disrupted, giving this biased result. Another small
contribution to this bin is from the sub-substructure. As seen in
Figure ~\ref{spiderpanel} the systems of substructure are spiraling
into the dark matter halo, thus the sub-structure is being destroyed
in its very own (sub-)host before completing a full orbit. Regardless
of this ``zero orbit'' peak, most of the disrupted satellites complete
at least the same number of orbits as the ones which survive to $z=0$.

\begin{figure}
  \centerline{\psfig{file=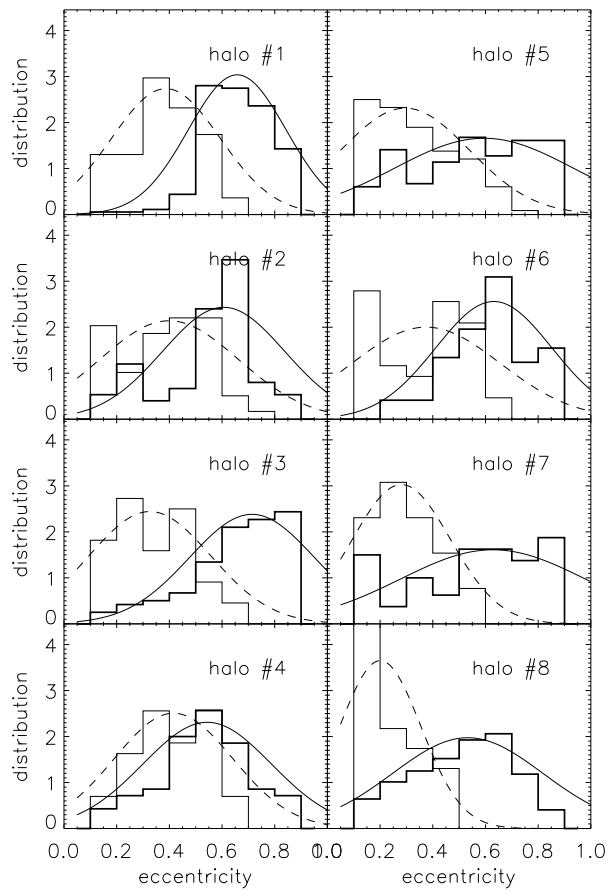,width=\hsize}}
  \caption{The orbital eccentricity distribution 
           function presented as histograms for all satellites that survived 
           until redshift $z=0$ (thick lines) and those which were destroyed
           (thin lines). The curves represent best-fit Gaussians to the distributions.}  
  \label{Necc}
\end{figure}

\begin{figure}
  \centerline{\psfig{file=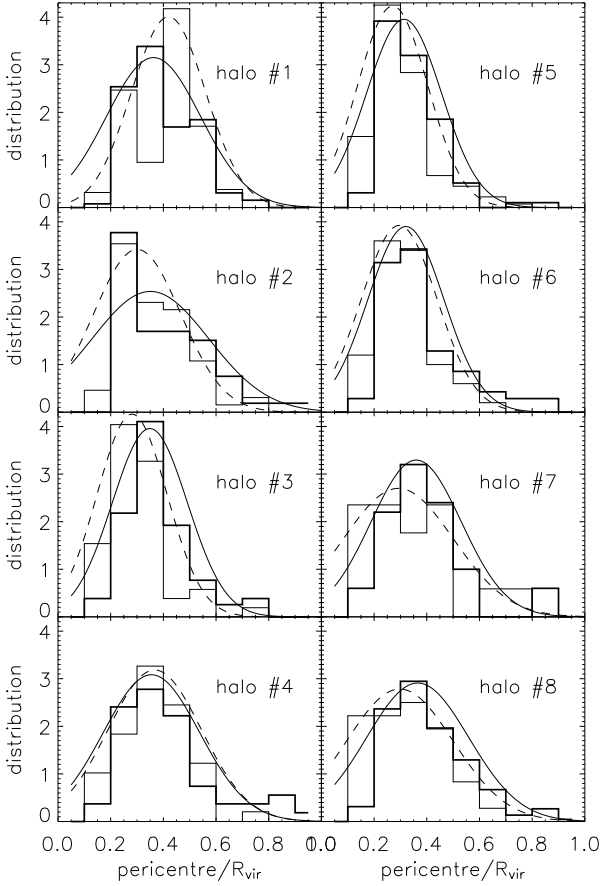,width=\hsize}}
  \caption{The pericentre distribution 
           function presented as histograms for all satellites that survived 
           until redshift $z=0$ (thick lines) and those which were destroyed
           (thin lines). The curves represent best-fit Gaussians to the distributions.}  
 \label{Nperi}
\end{figure}

\subsection{Orbital Eccentricity and Pericentres} \label{satEccPeri}

As outlined in \GKGI, we define eccentricity as

\begin{equation} \label{ecc}
e = 1 - \frac{p}{a} \ ,
\end{equation}

\noindent
where the pericentre $p$ and apocentre $a$ distances were those last 
measured in the
satellite's (or the tracer particle's) orbit. Using this definition we
show in \Fig{Necc} the distribution of the orbital eccentricity $e$ for
both live (thick lines) and disrupted satellites (thin lines).  Further,
we fit a Gaussian

\begin{equation} \label{Gaussian}
 P(e) = \frac{1}{\sqrt{2\pi\sigma^2}} \ e^{-\frac{(e-e_0)^2}{2\sigma^2}}
\end{equation}

\noindent
to the distributions (solid curves for live and dashed lines for
disrupted satellites). From this fit the peak orbital eccentricity for
the eight halos has an average of $\langle{e_0}\rangle=0.61$ with an
average standard deviation of $\langle{\sigma}\rangle=0.19$.  We do
not observe again any correlation of peak position and width of the
distribution with environment, age, host history or richness, even
though $\sigma$ appears to be larger for the younger dark matter
halos. It is interesting to note that if we stack the data from the
eight halos together and then separate the satellite population into
differing mass bins we see little variation. This common behaviour
still holds and probably reflects the scale free force of gravity and
hierarchical construction. The disrupted satellites, however, have a
different distribution, almost mirroring the survivors about the
$e=0.5$-axes, with the peak eccentricity near
$\langle{e^d_0}\rangle=0.34$ and a dispersion of
$\langle{\sigma^d}\rangle=0.16$. Thus, the destroyed satellites were
preferentially on more circular orbits.

The pericentre distributions provide additional insight into the
nature of the orbits of the disrupted satellites. The result for
redshift $z=0$ can be seen in \Fig{Nperi} where $p$ has been
normalised by the virial radius of the host. Again, live satellites
are represented as thick lines and disrupted ones as thin lines. A
striking characteristic of these distributions is again the similarity
between the halos. Moreover, we also observe a similarity in the
distribution for the live and disrupted satellites. This is emphasised
particularly by the best-fit Gaussians to the distributions. The
maximum peak lies at 35\% of the virial radius for live satellites
with a mean dispersion of $\langle\sigma\rangle=0.12$ as opposed to
31\% of \Rvir\ for the disrupted ones with a dispersion of
$\langle\sigma^d\rangle=0.11$.  Unlike eccentricity, the pericentre
distribution rises quickly towards the peak and falls off moderately
to the outer parts of the host halo. We already noted the lack of
correlation with mass, age, environment and richness, but there is,
however, a mild dependence on the state (live or dead) of the
satellite as disrupted satellites appear to have had marginally nearer
excursions towards the host centre. When we stack the data from the
eight halos and separate into differing satellite mass bins we again
see little variation.

In summary the difference between live and disrupted satellites lies
primarily in the eccentricity distribution. Disrupted satellites seem
to be on more circular orbits. Since the disrupted satellites have
similar pericentres to those which survive their circular nature means
that they spend more time in the deeper regions of the potential
well. Hence, they experience stronger tidal forces for longer periods,
and are thus disrupted/dissolved more readily.

\begin{figure}
 \centerline{\psfig{file=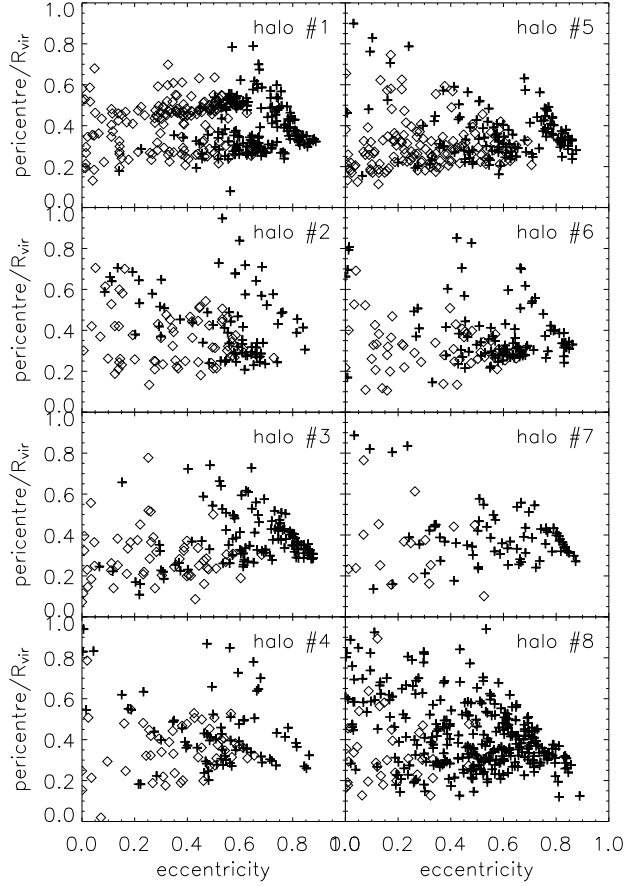,width=\hsize}}
 \caption{We plot the satellite eccentricity against pericentre normalised
          by the virial radius of the host. The crosses represent the satellites
          that survived until $z=0$ while the diamonds represent the disrupted 
          satellites.}
 \label{eccperi}
\end{figure}

We close this subsection with an examination of the
pericentre-eccentricity relation (\Fig{eccperi}). The crosses
represent the satellites that survived until redshift $z=0$, while the
diamonds represent the disrupted satellites.  Although many of the
features seen here are also seen in \Fig{Necc} and
\Fig{Nperi}, there are two additional ones which we will comment on now.

One such feature is the bimodality in the satellite distribution of
halo \# 1. In this Figure there seems to be two distinct (live and
dead) satellite populations orbiting within the dark matter halo. As
we saw, particularly in \Fig{spiderpanel}, this dark matter halo lies
in a rich area being fed by (at least) two filaments. Perhaps these
two satellite populations are a remnant of the filamentary large-scale
structure surrounding the host? If that is true, it is interesting
that the satellites still maintain their dynamical distinctiveness
after several orbits.

In addition there appears to be a distinct population of satellites with
an extremely tight
\mbox{(anti-)correlation} between eccentricities $e \approx 0.6-0.9$ and
pericentres $p\approx 0.3-0.8$ in many halos. Upon detailed
investigation these satellites were identified to have completed one
orbit and were entering the halo for the first time. This is
consistent with the notion of the satellites infalling on radial
orbits. We investigate this point in the next section, namely the
evolution of eccentricity.

\subsection{Evolution of Eccentricity}

\begin{figure}
 \centerline{\psfig{file=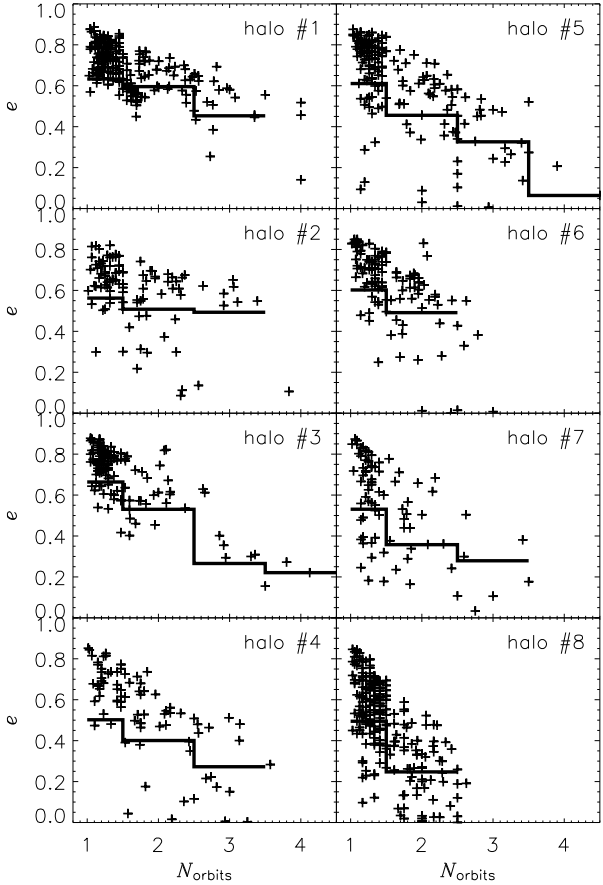,width=\hsize}}
 \caption{The final orbital eccentricity  plotted against the number of 
          orbits completed for each surviving satellite (crosses). The 
          histogram shows the averaged eccentricity (see text for details).}
 \label{eccOrbits}
\end{figure}

In the previous section we considered the orbital characteristics of
the satellites at redshift $z=0$. We now investigate the evolution of
the orbital eccentricities of the satellite family.

\Fig{eccOrbits} shows the eccentricity of each satellite
(represented by the crosses) versus the number of orbits the satellite
has completed. There is a clear trend for eccentricity to decrease as
the number of orbits of a satellite increases. This is also
demonstrated by the histogram, which is not the ``binned'' results of
the crosses, but is instead the average eccentricity for all
satellites that had $\geq N_{\rm orbits}$ orbits

\begin{equation}
 \langle e \rangle (N_{\rm orbits}) = \frac{1}{N_{\rm sat}({\ge N_{\rm orbits}})}
                    \sum_{i=1}^{N_{\rm sat}({\ge N_{\rm orbits}})} 
                    e_i(N_{\rm orbits}) \ .
\end{equation}

\noindent
Here, $e_i(N_{\rm orbits})$ is the eccentricity of satellite $i$ after
$N_{\rm orbits}$ orbits and $N_{\rm sat}({\ge N_{\rm orbits}})$ is the
number of satellites with equal or more than $N_{\rm orbits}$ orbits.

For example, a satellite that has had 3 orbits contributes its
respective value of eccentricity to the average eccentricity in the
bins for 1, 2, and 3 orbits. This histogram shows a trend indicative
of orbit circularisation with time. Before the work of Hashimoto \ea
(2002) one would have been quick to interpret this result as dynamical
friction circularising the orbits, however, they suggested
otherwise. To confirm this, we selected the satellites at differing
pericentre, as dynamical friction is proportional to the local density
of the background field, and thus has it's strongest influence at
pericentre. Having done this, we saw no significant change in the
above trend. Further, when we used the analytical predictions of
Taffoni \ea (2003) we found that very few satellites in the population
presented could be affected by dynamical friction. Therefore, we do
not attribute the circularising of the orbits to dynamical
friction. One mechanism which could be responsible for the
circularising of the orbits is the secular growth in the host halo's
mass. In response to this increase in host mass the velocity and hence
the orbit of the satellite changes. Further we suggest that this
change acts to circularise the orbit. Within the context of a fully
self-consistent \nbody\ simulation this claim is difficult to verify.
However, in \Sec{velbias} we show that a clear relationship exists
between satellite velocity dispersion and host halo mass, implying
that the satellite velocities respond to a change in host mass.

Finally we compare our orbital parameters to those presented by
Ghigna~\ea (1998).  The major difference between our respective
analysis in the derivation of the orbital characteristics is that even
though their host halos and satellite galaxies formed fully
self-consistently in cosmological simulations they took the satellite
positions and velocities and evolved them in a
\textit{static spherical potential} in order to obtain orbital
characteristics. Ghigna~\ea found that the satellites on radial orbits
were more likely to be disrupted than those on circular orbits because
they penetrate further into the dark matter halo potential well.  In
addition, radial orbits were quite common and circular orbits quite
rare in their simulations.  The average ratio of apocentre:pericentre
in their outputs was 6:1, with nearly 25\% of the halos on orbits with
ratios in excess of 10:1. In our language this 6:1 ratio equates to an
eccentricity of $\sim 0.83$. Hence, the Ghigna~\ea satellite orbits
were considerably more radial than ours. We can not reconcile this
discrepancy through dynamical friction arguments. However, it could
potentially be explained by the lack of a live halo. This explanation
correlates with our previous finding in
\Sec{SatAnalysis} for a recently infalling satellite
population. This latter population constitute a distinct subpopulation
with a strong
\mbox{(anti-)correlation} in \Fig{eccperi} and an average eccentricity
of $\sim 0.8$, quite similar to the result of  Ghigna~\ea.

\subsection{Circularity of Orbits}
Another way to measure the circularity of the orbits can be written 

\begin{figure}
 \centerline{\psfig{file=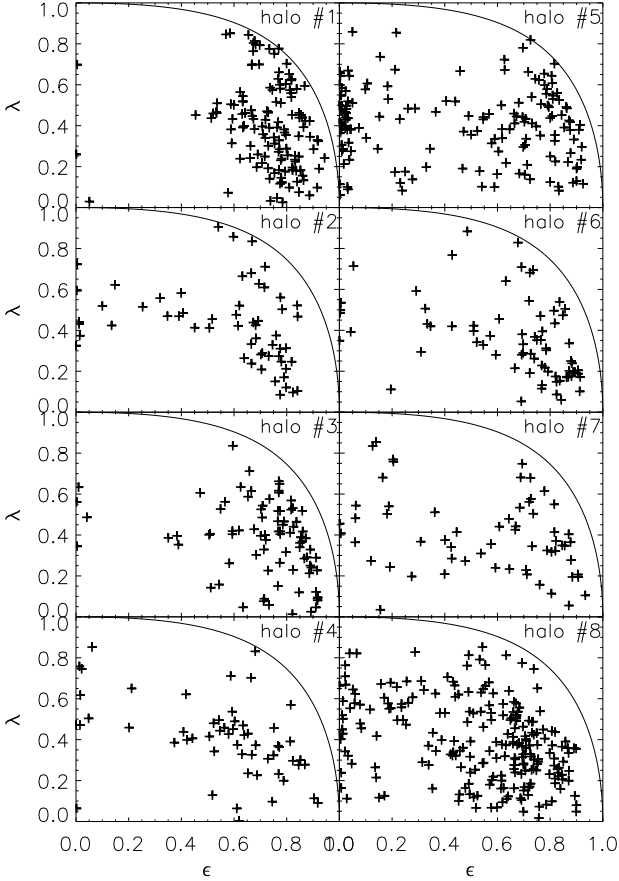,width=\hsize}}
 \caption{Circularity as a function of orbital eccentricity. Only
          satellites with at least one full orbit are considered
          and the solid line marks the 1:1 correlation.}
 \label{circularity}
\end{figure}

\begin{equation} \label{ldef}
 \lambda = \frac{J_{\rm sat(E)}}{J_{\rm circ(E)}}
\end{equation}

\noindent
where $\lambda$ is the ratio between the actual angular momentum of the
satellite $J_{\rm sat}$ and the angular momentum of a circular orbit
$J_{\rm circ}$ with the same energy $E$.

The correlation of $\lambda$ with the the orbital eccentricity defined
in \Eq{ecc} is presented in \Fig{circularity} for all satellites that
had at least one full orbit at redshift $z=0$. The solid lines are
upper limits given by the assumption that both satellite and host halo
are point masses. For a derivation of this relation please refer to
the Appendix.

We observe that most of the satellites follow the trend indicated by
the analytical estimate showing that low-eccentricity satellites are
in fact on more circular orbits.  However, the trend is only
suggestive but not strong with a significant scatter making it
difficult to substitute one measure for the other.


\section{Dark matter and satellite Kinematics}\label{velbias}

\begin{figure}
 \centerline{\psfig{file=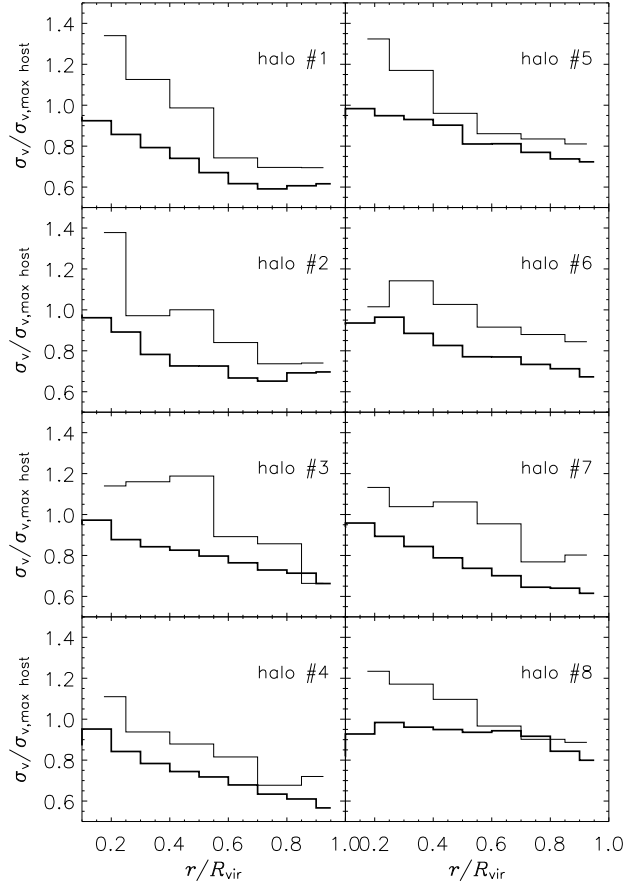,width=\hsize}}
 \caption{The normalised three dimensional velocity dispersion as a function 
          of halo-centric distance normalised by the maximum circular velocity 
          of the host halo. The dark lines represents the dark 
          matter distribution, while the light line is the satellite 
          distribution.}
 \label{vdisprad}
\end{figure}

A controversy still exists as to whether or not cluster members trace
the dark matter velocity distribution. Colin, Klypin~\& Kravtsov
(2000) found a substantial positive velocity bias when calculating the
local (or one-point) velocity bias

\begin{equation} \label{bv}
 b_v(r) = \frac{\sigma_{v,\rm sat}(r)}{\sigma_{v,\rm DM}(r)} \ .
\end{equation}

A similar yet less pronounced signal was reported by Okamoto~\& Habe
(1999), whereas Ghigna~\ea (1998) did not find any significant
velocity bias in their analysis of cluster substructure. Ghigna~\ea
(2000) re-visited this issue, obtaining a small positive bias, but
much weaker than that reported by Colin~\ea (1999). Springel~\ea
(2001) also examined this question, suggesting a small anti-bias from
their cluster simulations.

We now attempt to shed light on this controversy, by comparing the
kinematical properties of the host halo and the satellite
populations. We first look at the three-dimensional velocity
dispersion normalised by the maximum circular velocity of the host as
a function of (normalised) halo centric distance.  The results for all
eight host halos is presented in \Fig{vdisprad}. The dark matter
distribution of the halo is represented by the thick histograms while
curves based upon the satellite galaxy population are plotted as thin
histograms. Note that we do not see the characteristic ``rise and
fall'' for the velocity dispersion of a Navarro, Frenk~\& White (1997)
profile (cf.  Lokas~\& Mamon 2001), because for the host halos under
consideration $\sigma_v$ peaks at about 10\% of \Rvir, which is where
we start to plot the data.  In general though, from the inner bin at
10\% of \Rvir\ out to
\Rvir\ we see a drop in the dark matter's three-dimensional velocity
dispersion. This drop is even more pronounced for the satellite
population.  Thus in general for each halo we see an increasing
satellite ``local velocity bias'' as we get closer to the center of
the host halo. In the outer regions we have $b_v(\leq\!R_{\rm vir})
\sim 1.0$ while in the inner regions $b_v(\geq\!0.1R_{\rm vir})$
varies from $1.1-1.4$ with the largest value for $b_v$ being recorded
for halo~\#2, $b_v(\approx\!0.1R_{\rm vir}) \sim 1.4$. Essentially for
all halos though $b_v(r)
\geq 1.0$ at all radii.  This result agrees with the data shown by
Colin~\ea (1999) and Ghigna~\ea (2000), in which both groups found
that the satellite population is ``positively'' biased with
$b_v(\approx\!0.1R_{\rm vir}) \sim 1.2-1.3$. However, it contrasts
with the Springel~\ea (2001) findings in which a small negative
velocity bias for the central regions of the cluster was claimed.

There is no discernible relationship between the eccentricity
distribution (cf. \Fig{Necc}) and the bias seen in \Fig{vdisprad};
however, there is sufficient orbital distribution variation to
accommodate for the variation in the bias. Improved statistics will be
needed to decouple the detailed variations. However, we have shown
that the velocity bias has the same basic shape for all eight halos,
strengthening the case that this is a general characteristic of the
satellite population.

\begin{table}
\caption{The global velocity bias for the eight host dark matter halos, within the virial radius.}
 \label{velbiastable}
\begin{center}
\begin{tabular}{ccccc}\hline
halo & $\sigma_{v,halo}$ & $\sigma_{v,sats}$ & $b_{v, \rm global}$  & $b_{v, \rm global}^{N_{\rm orbits}\ge1}$ \\
\hline \hline
 \#       1  &    1140  &   1077  &  1.12 &  1.01 \\
 \#       2  &    887   &   811   &  1.07 &  1.01 \\
 \#       3  &    900   &   834   &  1.09 &  0.98 \\
 \#       4  &    789   &   729   &  1.06 &  0.99 \\
 \#       5  &    1170  &   1168  &  1.08 &  0.98 \\
 \#       6  &    857   &   807   &  1.12 &  1.03 \\
 \#       7  &    811   &   852   &  1.19 &  1.13 \\
 \#       8  &    1067  &   1119  &  1.09 &  1.02 \\
\end{tabular}
\end{center}
\end{table}

Using all the dark matter particles and satellite galaxies within the
virial radius we also calculate the global velocity bias

\begin{equation}
 b_{v, \rm global} =  \frac{\sigma_{v,\rm sat}(<\!R_{\rm vir})}
                           {\sigma_{v,\rm DM}(<\!R_{\rm vir})}
\end{equation}

\noindent
which is summarized for all eight halos in
Table~\ref{velbiastable}. It is interesting to note that we find a
slight positive bias, even though Klypin~\ea (1999) and Ghigna~\ea
(1998) both claimed that such a global velocity bias should not
exist. However, Klypin \ea (1999) did hint that the possibility of a
mild negative bias might exist, based upon earlier theoretical work by
Carlberg (1994). Carlberg argued that as galaxies fall into the
cluster for the first time they lose energy to the cluster and become
systematically more bound, orbiting freely as coherent,
self-gravitating units. From his simulation he measured a value of
$b_{v, \rm global}
\sim 0.8 \pm 0.1$.  From our halos we find an average global velocity bias
$\langle b_{v, \rm global}\rangle \sim 1.103 \pm 0.002$, which is only
a very slight bias. If we restrict the global velocity bias
calculation to satellites that have had at least one orbit, $\langle
b_{v,
\rm global}^{N_{\rm orbits}\ge1}\rangle \sim 1.019 \pm 0.002$, a 7\%
decrease in the bias. The quoted errors are the scatter within the eight halos.
  
\begin{figure}
 \centerline{\psfig{file=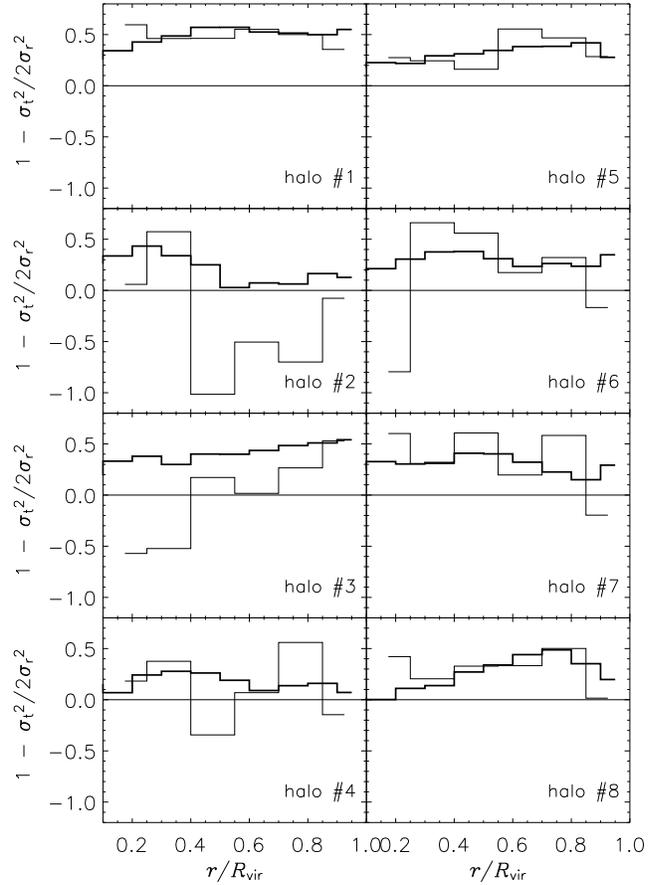,width=\hsize}}
 \caption{The velocity anisotropy parameter as a function of halocentric distance. The thick line represents the dark matter distribution, while the thin line is the satellite distribution. The quantities were calculated in linear radial bins.}
 \label{betarad}
\end{figure}

To further investigate the velocity bias we measure the variation from
an isotropic velocity distribution using the anisotropy parameter, as
seen in Figure ~\ref{betarad}

\begin{equation}
 \beta = 1 - \sigma_{t}^{2}/2\sigma_{r}^{2} \ , 
\end{equation}

\noindent
where $\sigma_t$ measures the variance in the tangential and
$\sigma_r$ in radial direction. Once again, the dark matter
($\beta_{\rm DM}$) and satellite distribution ($\beta_{\rm sat}$) were
averaged in linearly spaced radial bins, normalised by the virial
radius. The dark matter distribution is represented by the thick
histogram and the satellite galaxies by the thin histogram.  For an
isotropic distribution $\beta = 0$; if $\beta \rightarrow 1$,
then the velocities are preferentially radial; if $\beta
\rightarrow -\infty$, they are preferentially tangential.

The standard way to represent $\beta$ is on a logarithm scale starting
from the inner percent of the virial radius.  Cole \& Lacey (1996) and
Thomas \ea (1998) employed this technique and measured a linear
increase in $\beta_{\rm DM}$, starting with essentially isotropic
orbits close to the halo core and becoming increasingly radial in the
outer regions.  In keeping with previous figures we plot the data on a
linear scale.  This makes it difficult to compare to this
work. However, our value for $\beta_{\rm DM}$ does follow the same
general radial relationship seen in earlier studies.

In general $\beta_{\rm sat}$ for the substructure population (broadly)
follows the dark matter distribution.  However, for halos~\#2 and \#3,
satellite orbits are generally more tangential than the dark matter
background.

Finally, to gain further insight into the velocity distributions of
the satellite and dark matter distributions we investigated the radial
variation of the radial velocity component. Note we do not present a
figure for this results.  When normalised by the maximum circular
velocity for all eight halos, the fraction $v_{r, \rm DM}/V^{\rm
max}_{\rm circ}$ is essentially zero. Although, we do see
signs of a slight dip in the outer regions. This dip is a
characteristic signature of infall into the system.

\subsection{Observational Impact}\label{velobs}

\begin{figure}
\centerline{\psfig{file=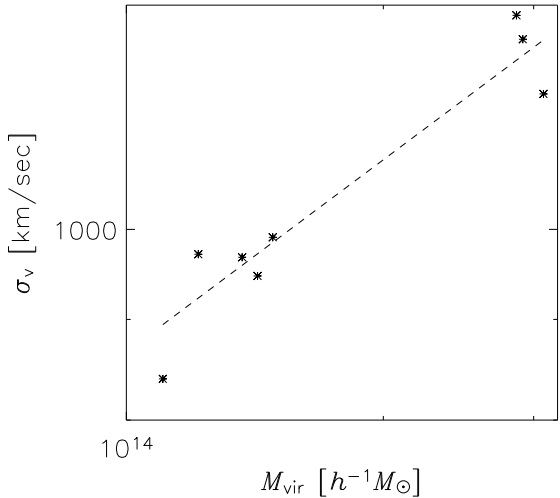,width=\hsize}}
 \caption{The velocity dispersion of the satellite galaxies within the virial radius of the host halo as a function of host halo mass.}
 \label{dismass}
\end{figure}

Astronomers traditionally determine the mass of galaxy cluster by
assuming that the galaxy distribution is virialised.  The virial
theorem then gives (Heisler, Tremaine \& Bahcall 1985):

\begin{equation} \label{obmass}
 M_{\rm vir} = \frac{R_{\rm vir} \sigma_v^2}{G} \ .
\end{equation}
 
In practice, however, the observational determination of the correct
radius \Rvir\ and velocity dispersion $\sigma_v$ are far from
straightforward, as it is difficult to determine whether or not a
galaxy is a member of a cluster.  Furthermore, even though we have
just learned that the satellites are ``stationary'', the existence of
the slight global velocity bias (cf. \Table{velbiastable}) does impact
the determination of cluster masses. Or in other words, the observed
positive bias will result in an over-estimation of the halo mass when
relying on the dynamics of its substructure.

From our simulations we find (to no surprise) that the velocity
dispersion of a system of galaxies is a reliable indicator of the
depth of the potential associated with the system. Even though we do
not have a broad mass range ($M_{\rm vir}\sim 1-3 \times
10^{14}$\hMsun) to test this hypothesis we reconstruct the expected
relationship between mass
\Mvir\ and velocity dispersion $\sigma_{v, \rm sat}$ of the satellite
galaxies within the virial radius of the host halo. The results are
presented in \Fig{dismass}.  Each of our halos scatter about the
expected analytical relationship:

\begin{equation}
 \sigma_v \sim M^{1/3}_{\rm vir}
\end{equation}

\noindent
This relationship can easily be understood analytically when we combine
\Eq{obmass} and our definition for virial radius $\Delta_{\rm vir} \rho_b
= 3 M_{\rm vir}/(4 \pi R^3_{\rm vir})$. The solid line displayed in
the figure is the best fit power law to our eight data points. The
logarithmic slope was found to be $\sim 0.329$, sufficiently close to
the expected value of $1/3$ .

\section{Conclusions} \label{Conclusions}

If the hierarchical model of structure formation is correct then the
dynamics of satellite galaxies are an important ingredient to also
understanding the formation and the evolution of galaxies. Therefore,
in this paper we presented a series of self-consistent cosmological
simulations of dark matter halos with the required mass and spatial
resolution to follow satellite galaxies orbiting even within the
central regions of the host potential. Moreover, the simulations had
sufficient time resolution to actually resolve the satellite dynamics
with high precision.

The first part of this study was dedicated to analysing and describing
the differences in the host halos.  These host halos were chosen to
sample a variety of triaxialities, formation times and mass/satellite
accretion, despite being of comparable mass. The halos also had very
different formation histories, from quiescent to violent. When
investigating the halo environment we quantified a value of
``richness'' which was defined to be the fraction of satellite
galaxies that the halo had accreted since its formation. One
interesting result from this analysis was the similar rate of
satellite disruption seen for all eight halos. Furthermore, over the
history of each of our halos on average about 10\% of its respective
mass is locked up in the satellite galaxies.  Much of this mass can be
attributed to massive satellites rapidly captured by dynamical
friction.

But even though the eight dark matter host halos were quite different,
their respective satellite population showed remarkable similarities.
The average orbital eccentricity of the satellites was found to be $e
\approx$ 0.61 with minimal scatter ($\sigma\approx0.19$). Moreover, the
average pericentre distance of the satellites was $p \sim$ 35\% of the
virial radius for all halos, again with minimal scatter
($\sigma\approx 0.12$). Satellites that were disrupted while orbiting
within the host's virial radius were replaced with a mass-less tracer
particle and hence we were also able to present their orbital
parameters at redshift $z=0$. We found that even though they have
smaller eccentricities ($e^d\approx 0.34$) than the surviving ones,
their pericentre distributions are nearly identical.  Since the
pericentres distributions of both surviving and disrupted satellites
were similar, implication is that the disrupted satellites spend more
time in the deeper regions of the potential well. As such, they
experience stronger tidal forces for longer periods, and are thus
being disrupted more readily. We also noticed that satellites with
more orbits tend to have smaller eccentricities. Difficult to explain
through the application of dynamical friction we attribute this to the
satellite's response to the growing host halo.

We also found that the local velocity bias at all radii is greater
than one and this increases as we move closer to the halo
centre. Since this is a characteristic for each of our halos, it
strengthens the case that this is a general pattern of the satellite
population in dark matter halos.  For the global velocity bias we find
an average $\langle b_{v, \rm global}\rangle \sim 1.103
\pm 0.002$, a  slight, but significant, positive bias. Further, if we
restrict the global velocity bias calculation to satellites that have
had at least one orbit we observe a 7\% decrease in bias to $\langle
b_{v, \rm global}\rangle
\sim 1.021 \pm 0.002$. Thus when we just
consider the ``virialised'' satellites, the bias nearly vanishes.
Finally we recovered the $\sigma_v \sim M^{1/3}$ relationship between
satellite velocity dispersion and halo mass.
  
Surprisingly, all the above stated results appear to be independent of
the actual host halo and its history. We were unable to identify any
trends with richness, triaxiality and/or formation time (other than
the number of orbits). Such similarities are suggestive of potential
additional underlying CDM universal laws.

\section{Acknowledgments} 
The simulations presented in this paper were carried out on the
Beowulf cluster at the Centre for Astrophysics~\& Supercomputing,
Swinburne University.  We also wish to thank Daisuke Kawata, Chris
Brook, and Rob Proctor for their helpful feedback throughout the
preparation of this manuscript. We acknowledge the financial support
of the Australian Research Council through its Federation Fellowship
(MAD) and Discovery Projects DP0208445 (MAD) and DP0343508 (BKG). MAD
also acknowledges the support of the Australian National University.

The useful advice provided by the referee Fabio Governato is
gratefully acknowledged, too.


\section*{Appendix: Analytical Estimate for Circularity Parameter}
We wish to establish a correlation between the circularity

\begin{equation}
 \lambda = \frac{J_{\rm sat}(E)}{J_{\rm circ}(E)} \ ,
\end{equation}

\noindent
where $J_{\rm sat}(E)$ measures the angular momentum of an individual
satellite with total energy $E$. $J_{\rm circ}(E)$ is the
corresponding angular momentum for a circular orbit with the same
energy $E$.

The connection needs to be derived using the fact that both
orbits share the same total energy, so we wish to express $E$ as
a function of $J$ in both cases, where the general expression for
$E$ and $J$ reads:

\begin{equation} \label{EL}
 \begin{array}{lcl}
  E & = & \frac{1}{2} v^2 - \frac{GM}{r}\\
\\
  J & = & |\vec{r} \times \vec{v}|\\
 \end{array}
\end{equation}

\subsection*{Circular Orbits}
For a circular orbit we know that $|\vec{r} \times \vec{v}|$ reduces
to 

\begin{equation}
 J_{\rm circ} = r v_{\rm circ}
\end{equation}

\noindent
as both vectors are always orthogonal. Moreover, the circular velocity
is given by

\begin{equation}
 v^2_{\rm circ} = \frac{GM}{r}
\end{equation}

\noindent
where $M$ is the total mass of the host halo (assumed to be a point
mass).

Combining this knowledge with \Eq{EL} we arrive at

\begin{equation}
 J_{\rm circ} = GM \frac{1}{\sqrt{-2E}}
\end{equation}

\subsection*{Keplerian Orbits}
For Keplerian orbits the situation is a little more complicated. We
know that the solution in spherical coordinates for closed orbits
looks like

\begin{equation} \label{Circu}
 r = \frac{J_{\rm sat}^2}{GM (1- \epsilon \cos{\phi})} \ ,
\end{equation}

\noindent
where $\epsilon$ is the measure for ellipticity of the orbit as given by 

\begin{equation}
 \epsilon = f/a
\end{equation}

\noindent
Here $f$ is the focal length and $a$ the major axis of the ellipse.
Geometry relates $\epsilon$ to our eccentricity $e$ as follows

\begin{equation} \label{eE}
\epsilon = \frac{2}{2-e} \ .
\end{equation}

We still need an expression for the velocity which can be derived by
using the first derivative of $\vec{r}$ with respect to time

\begin{equation}
 \frac{d\vec{r}}{dt} = \frac{d (r \vec{e}_r(\phi))}{dt} 
                     = \dot{r} \vec{e}_r + r \dot{\phi} \vec{e}_{\phi}
                     = \dot{r} \vec{e}_r + \frac{J_{\rm sat}}{r} \vec{e}_{\phi}
\end{equation}

Evaluating $\dot{r}$ and rearranging terms gives

\begin{equation}
 v^2 = \frac{J_{\rm sat}^2}{r^2} \frac{1-2\epsilon\cos{\phi} + e^2}{(1-\epsilon\cos{\phi})^2}
\end{equation}

Inserting this knowledge into \Eq{EL} again leads to

\begin{equation} \label{Kepler}
 E = \left(\frac{GM}{J_{\rm sat}}\right)^2 \frac{\epsilon^2-1}{2} \ .
\end{equation}

\subsection*{The Relation}
We know have everything necessary to write  the sought-after
relation between eccentricity $\epsilon$ and circularity $\lambda$;
we simply need to combine \Eq{Circu} and \Eq{Kepler}, keeping \Eq{eE} 
in mind:

\begin{equation}
 \begin{array}{lcl}
  \epsilon       & = & \displaystyle \frac{2}{2-e} \\
\\
  \lambda & = & \sqrt{1-\epsilon^2}\\
 \end{array}
\end{equation}

This relation is plotted as a solid line in \Fig{circularity}.

\end{document}